\documentclass[letterpaper, 10 pt, conference]{ieeeconf}  % Comment this line out if you need a4paper

\IEEEoverridecommandlockouts                              % This command is only needed if 
\usepackage[usenames, dvipsnames]{color}
\usepackage{xcolor}
\usepackage{amssymb}
\usepackage{mathtools}
\usepackage{enumerate}
\usepackage{amsmath}
\usepackage{array}
\usepackage{algorithm}
\usepackage{subcaption}
\usepackage{balance}
\usepackage[noend]{algpseudocode}
\usepackage{cite}
\usepackage[normalem]{ulem}
\useunder{\uline}{\ul}{}
\usepackage{pdfpages}
\makeatletter
\newcommand*{\rom}[1]{\expandafter\@slowromancap\romannumeral #1@}

\makeatother

\newtheorem{theorem}{Theorem}

\newtheorem{definition}{Definition}%[section]
\newtheorem{proposition}[theorem]{Proposition}%[section]

\newif\ifComments
\Commentsfalse
\usepackage{graphicx}
\graphicspath{{Images/}}
\DeclareGraphicsExtensions{.pdf,.emf,.png,.eps}

\title{\LARGE \bf
Grammar-based Representation and Identification of Dynamical Systems
}

\author{Dhruv Khandelwal, Maarten Schoukens and Roland T\'oth% \left \langle-this % stops a space
\thanks{This research is supported by the Dutch Organization for Scientific Research (NWO, domain TTW, grant: 13852) which is partly funded by the Ministry of Economic Affairs.}% \left \langle-this % stops a space
\thanks{All authors are with the Control Systems group, Department of Electrical Engineering, Eindhoven University of Technology, 5600 MB Eindhoven, The Netherlands.}%
\thanks{Corresponding author: {\tt\small D.Khandelwal@tue.nl}}
}

\begin{document}

\maketitle
\thispagestyle{empty}
\pagestyle{empty}

%%%%%%%%%%%%%%%%%%%%%%%%%%%%%%%%%%%%%%%%%%%%%%%%%%%%%%%%%%%%%%%%%%%%%%%%%%%%%%%%
\begin{abstract}
In this paper we propose a novel approach to identify dynamical systems. The method estimates the model structure and the parameters of the model simultaneously, automating the critical decisions involved in identification such as model structure and complexity selection. In order to solve the combined model structure and model parameter estimation problem, a new representation of dynamical systems is proposed. The proposed representation is based on Tree Adjoining Grammar, a formalism that was developed from linguistic considerations. Using the proposed representation, the identification problem can be interpreted as a multi-objective optimization problem and we propose a Evolutionary Algorithm-based approach to solve the problem. A benchmark example is used to demonstrate the proposed approach. The results were found to be comparable to that obtained by state-of-the-art non-linear system identification methods, without making use of knowledge of the system description.
\end{abstract}

%%%%%%%%%%%%%%%%%%%%%%%%%%%%%%%%%%%%%%%%%%%%%%%%%%%%%%%%%%%%%%%%%%%%%%%%%%%%%%%%
\section{INTRODUCTION}

\ifComments
\textcolor{red}{\begin{itemize}
\item Introduce the notion of model class, and its relevance in the context of system identification.
\item There exists a large variety of identification approaches for the different types of model classes. Most of these approaches cannot be applied successfully to other model classes.
\item Position the paper as a first step to develop a method that is capable of identifying models belonging to multiple model classes.
\item The final objective would be to develop an identification approach that not only identifies parameters in a model, but also the structure.
\item Introduce the analogy between a model and its model structure and a sentence and its grammatical structure. This analogy will the drive the motivation for using TAG representations.
\end{itemize}
}
\fi

The problem of inferring models from data has been well studied in many research domains including systems and control. Modelling of complex systems using first principle models can be too cumbersome to derive. 
%These difficulties arise due to a number of reasons such as:
%\begin{itemize}
%	\item Lack of knowledge of the true underlying laws that govern the behaviour,
%	\item Inability to extrapolate (interpolate) the known physical laws to a macroscopic (microscopic) level,
%	\item The complexity arising due to the interaction of multiple agents within the system.
%\end{itemize}
This led to the development of several grey-box and black-box modelling approaches \cite{Ljung1999}. 
%Over the years, within the systems and control community, many data-based identification techniques have been developed to address the unique challenges posed by various classes of systems such as linear, non-linear, time-varying, networked or distributed parameter systems \cite{ljung2015system}.
Each of these methods have been developed and tuned to identify a well specified class of dynamical systems described by the so-called \emph{model class} associated with the method \cite{Ljung1999}. However, they are, in general, not well-equipped to identify systems that do not belong to the assumed model class. As a result, most identification procedures require an expert practitioner to make several critical choices and ensure the validity of key assumptions, including model class, complexity and noise structure, to successfully complete a data-driven modelling task \cite{Ljung1999}.

Due to these critical user choices, even for a skilled practitioner, the task of modelling complex systems remains arduous, making automation of these choices highly challenging. % To address these challenges, black-box modelling approaches were developed. Such approaches make use of flexible, typically non-parametric models such as Gaussian processes or artificial neural networks \textcolor{red}{cite}. However, black-box models are typically less amenable for control or systems analysis applications. 
In this contribution, we develop a framework for system identification that can function across different model classes and automatically explores varying levels of complexity. In order to realize a method that can work across different classes of models, we propose a new representation for stochastic dynamical models. The proposed representation uses Tree Adjoining Grammar (TAG) \cite{joshi1987introduction} - a tree-generating system formulated for application in natural languages processing. The use of TAG-based representation allows one to express a model in terms of a set of fundamental building blocks called \emph{elementary trees}. These building blocks can be combined in specific ways in order to build more complex models. The advantage of using TAG-based representations is that a given set of elementary trees may be used to generate models that belong to different classes of dynamical systems. This allows the proposed identification framework to function across different model classes.% Incorporating the model structure in the representation is a stepping stone towards estimating model structure from data.

In the context of unknown model structure and complexity, the problem of inferring models from data can be divided into two components: i) the search of the \textit{appropriate} model structure and complexity, and ii) optimization of the model parameters. It is crucial to distinguish between these two problems since the former is a combinatorial optimization problem, while the latter is a continuous optimization problem, which may or may not be a convex problem, depending on the structure of the model \cite{Ljung1999}. %The use of TAG-based representations allows us to efficiently handle the involved combinatorial problem. 
In order to solve the combinatorial problem, we use an algorithm based on Genetic Programming (GP) and TAG. TAG makes the combinatorial search more efficient by restricting the ways in which the elementary trees can combine. Moreover, since the combinatorial search is formulated on the set of elementary trees of a TAG and \textit{not} a particular model structure, the GP-based algorithm for structure estimation runs independent of the model structure.

In order to effectively illustrate the idea, in this paper, we consider the problem of identifying models across model structures that belong to the superset of SISO (Single-Input Single-Output) polynomial NARX (Non-linear Auto-Regressive with eXogenous inputs) model class \cite{billings2013nonlinear}. The polynomial NARX class contains a number of commonly-used model structures such as FIR, ARX and truncated Volterra series. We propose a TAG for the class of polynomial NARX models. It will be shown that the TAG can be systematically scaled down to restrict the scope of the identification task. Similarly, it is also possible to extend the proposed TAG in order to broaden the scope. This task will be taken up in future research.% Furthermore, the user need not pre-specify the maximum complexity of the models, either in terms of maximum lags or maximum exponents of the monomials. The proposed identification procedure identifies models across varying levels of model complexity, thereby resulting in a pareto-set of identified models.
% This shifts the focus from the utilization of a model for the user's benefit to the computation of the model itself, a task that is typically quite expensive for most industries. Moreover, the analysis usually imposes restrictions or simplifications on the underlying system to make the problem tractable. While these assumptions may be valid for a big class of systems, they are not always applicable in general.

%Thus, in this paper, we present an approach for identification of models within the polynomial NARX class that does not rely on the user for structural or complexity-related information of the system.

%\section{Overview of Evolutionary Algorithms and Grammar-based methods in SI}
\ifComments
\textcolor{red}{
\begin{itemize}
\item The literature review is currently focused on all the Evolutionary Algorithm-based approaches in the literature for structure detection in system identification.
\item The review should probably also include non-parametric approaches, since they circumvent the problem by using (possibly infinite) basis functions, and control complexity by regularization. 
\end{itemize}
}
\fi
The idea to use Evolutionary Algorithms (EAs) for model structure selection in system identification (SI) is not new. In \cite{fonseca1996non} the authors used Multi-Objective Genetic Algorithms (MOGA) to perform term-selection in polynomial NARMAX models. This approach was extended to rational NARMAX models in \cite{rodriguez2000use}. A combination of Genetic Programming (GP) and Orthogonal Least Squares (OLS) was used to identify non-linear input-output models (such as polynomial ARMA and truncated Volterra series) in \cite{madar2005genetic}. Along similar lines, the authors in \cite{quade2016prediction} us multi-objective GP to estimate dynamical models in a structure-free manner. In \cite{kristinsson1992system}, the authors use Genetic Algorithm (GA) to estimate the pole-zero locations of ARMAX models. The key idea in these methods is similar - EA is used to explore models of varying complexities within a pre-determined model structure. However, the chosen model structure determines the encoding of the model into a representation that is appropriate for EA. This makes it difficult to automate the aforementioned approaches for different model structures. This is also exemplified in \cite{worden2018evolutionary}, where multiple EA-based approaches are developed in order to identify different benchmark examples.% For these benchmark problems, the results found in \cite{worden2018evolutionary} are some of the best found in the literature.% The work in \cite{worden2018evolutionary} also exemplifies the need of physical insight in non-linear SI.

In contrast to the existing literature, our proposed EA-based SI method uses TAG-based representations to separate the numerical tools (for combinatorial optimization) from the choice of model class. Different model classes can be represented using TAG, thereby ensuring that the same EA-based approach can be used for any user-defined TAG. By isolating the numerical methods for structure detection from the model class chosen, we obtain an approach that can be easily scaled down to smaller sub-classes of models or scaled up to larger model classes. This is an essential development needed to automate the task of model structure and complexity selection.

Previous attempts to use TAG in a data-driven modelling context has been made in \cite{hoai2002solving} to estimate static models for a curve-fitting problem. However, the extension of the idea to dynamical systems, as proposed in this paper, is far from trivial. For instance, the presence of noise in the data may introduce bias in the estimates if the noise contributions are not treated carefully. In this paper, we extend TAG representation to stochastic dynamical models.

\section{Tree Adjoining Grammar of Dynamical Models}

\ifComments
\textcolor{red}{
\begin{itemize}
\item Introduce TAG. Give bare minimum descriptions of all the notions that will be needed later.
\item Propose TAG for NARX model structure. Only provide sketch of the proof.
\item Discuss some aspects of the proposed TAG - the new TAG-based representations of the NARX model, other model structures included in the formulation,...
\end{itemize}
}
\fi

\subsection{Preliminaries}

TAG \cite{joshi1987introduction} is a tree generating system that was developed from natural linguistic considerations. TAGs were initially developed in order to capture features of natural languages that could not be captured by Context Free Grammars (CFG). Several formal properties can be attributed to TAG and can be found in \cite{joshi1997tree}. To develop our contribution, we briefly introduce some key ingredients of TAG. For a more detailed treatment of TAG, see \cite{joshi1997tree} and \cite{kallmeyer2009declarative}.

\begin{definition}[Finite tree]
	A \textit{finite tree} is a directed graph, denoted by $\gamma =\left \langle V,E,r\right \rangle$, where $V$ is the set of vertices, $E$ is the set of edges, and $r \in V$ is the root node and $\gamma$ has the following properties:
		\begin{itemize}
		\item[-] $\gamma$ contains no cycles,
		\item[-] Only $r \in V$ has in-degree 0 (i.e., number of incoming edges),
		\item[-] Every vertex $v \in V$ is accessible from $r$,
		\item[-] All nodes $v \in V \setminus \{r\}$ have in-degree 1,
		\item[-] A vertex with out-degree 0 (i.e., number of outgoing edges) is a leaf. 
		\end{itemize}
\end{definition}
Introduce the labelling function of a tree $l:V \rightarrow A$ that maps from the set of vertices $V$ of a tree to an alphabet (i.e., set of symbols) $A$.
\begin{definition}[Tree Adjoining Grammar]
	A Tree Adjoining Grammar $G$ is a tuple $\left \langle N,T,S,I,A \right \rangle$, where
	\begin{itemize}
		\item[-] $N$ is an alphabet of non-terminal symbols;
		\item[-] $T$ is an alphabet of terminal symbols;
		\item[-] $S$ is a specific start symbol in $N$;
		\item[-] $I$ is a set of {initial trees}. An initial tree $\gamma = \left \langle V,E,r\right \rangle$ has $l(v_\mathrm{int}) \in N$ for all internal vertices $v_\mathrm{int} \in V$ and $l(v_\mathrm{leaf}) \in (N \cup T) \setminus \{l(r)\}$ for all leaves $v_\mathrm{leaf} \in V$;
		\item[-] $A$ is a set of \emph{auxiliary trees}. A tree $\gamma = \left \langle V,E,r\right \rangle$ is an auxiliary tree iff $l(v_\mathrm{int}) \in N$ for all internal vertices $v_\mathrm{int} \in V$ and $l(v_\mathrm{leaf}) \in (N \cup T)$ for all leaves $v_\mathrm{leaf} \in V$ and there is a unique leaf $f \in V$ with $l(f) = l(r)$. The node $f$ is called the \emph{foot node}. The auxiliary tree is denoted as $\left \langle V,E,r,f\right \rangle$.
	\end{itemize}
\end{definition}

% A TAG $G$ can be defined as a 5-tuple consisting of:
%\begin{itemize}
%	\item Two disjoint alphabets (i.e., sets of symbols) $N$ and $T$ that contain \emph{non-terminal} and \emph{terminal} symbols respectively. Non-terminal and terminal symbols are the labels of internal nodes and leaves, respectively, of all the trees that can be generated from a TAG.
%	\item A set of \emph{initial trees} $I$ and a set of \emph{auxiliary trees} $A$.  An initial tree is a tree in which all internal nodes have labels belonging to $N$, while all leaves have labels belonging to $N \cup T$ except the label of the root node of the initial tree. An auxiliary tree is a tree in which all internal nodes have labels belonging to $N$, while all leaves have labels belonging to $N \cup T$, such that at least one of the leaves has the same label as the root of the initial tree. Initial and auxiliary trees, collectively known and \emph{elementary trees}, capture the the structural rules for re-writing of a tree using the given TAG.
%	\item A start symbol $S \in N$.
%\end{itemize}

The set if trees $I \cup A$ is called \emph{elementary trees}. A \emph{syntactic tree} is a tree $\gamma = \left \langle V,E,r\right \rangle$ that satisfies $l(r) \in S$, $l(v_\mathrm{int}) \in N$ for all internal vertices $v_\mathrm{int} \in V$ and $l(v_\mathrm{leaf}) \in (N \cup T)$ for all leaves $v_\mathrm{leaf} \in V$. A syntactic tree is said to be \emph{saturated} if all leaves $v_\mathrm{leaf} \in V$ satisfy $l(v_\mathrm{leaf}) \in T$. The set of all finite saturated trees of a TAG is called the \emph{tree language} $L_\mathrm{T}(G)$ of the TAG $G$. The \emph{yield} of a saturated tree is the string of labels of the leaves of the tree. The set of yields of the trees in $L_\mathrm{T}(G)$ is called the string language $L(G)$.

\begin{figure}
		\vspace*{0.2cm}
		\centering
		\begin{subfigure}[t]{0.9\linewidth}
			\includegraphics[scale = 0.41]{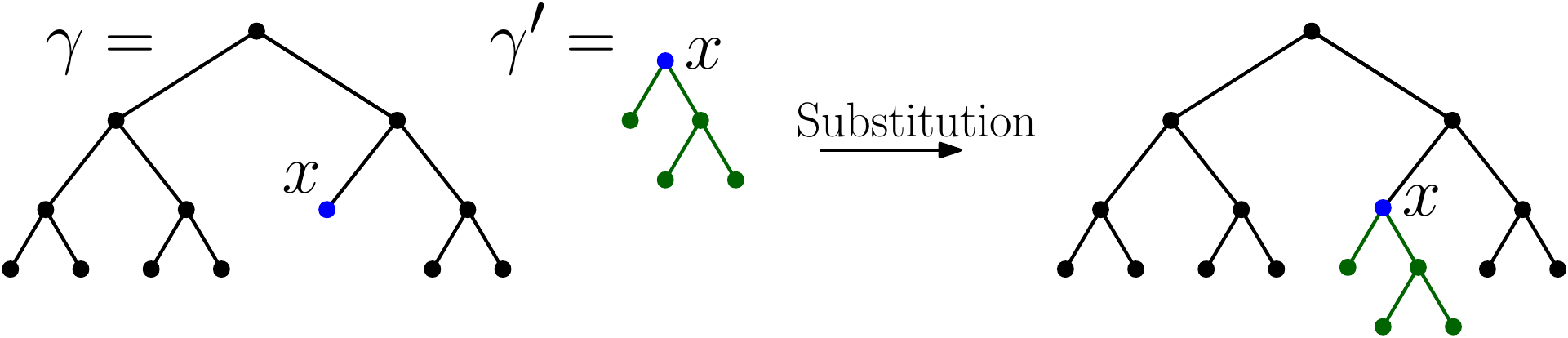}
			\caption{TAG substitution operation.}
			\label{fig:substitution}
		\end{subfigure}
		\\
		\begin{subfigure}[t]{0.9\linewidth}
			\includegraphics[scale = 0.41]{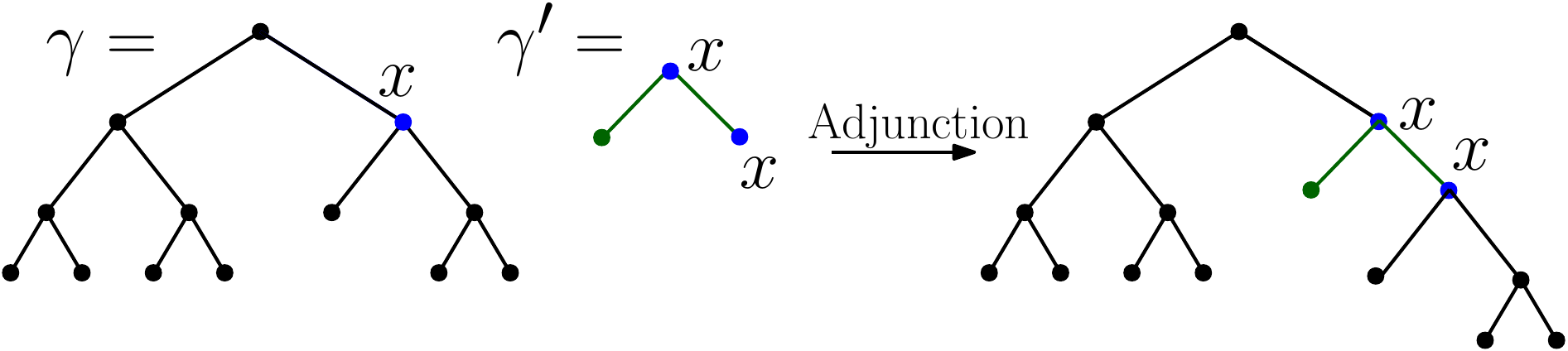}
			\caption{TAG adjunction operation.}
			\label{fig:adjunction}
		\end{subfigure}
		\caption{Illustration of the TAG operations.}
		\label{fig:operations}
\end{figure}

The TAG framework provides two operations, \emph{substitution} and \emph{adjunction}, that can be used to generate or re-writing tree structures from a TAG $G$.
\begin{itemize}
	\item[-] Substitution allows one to substitute a leaf $v_\mathrm{leaf}$ in a syntactic tree $\gamma = \left \langle V,E,r\right \rangle$ with an initial tree $\gamma' = \left \langle V',E',r'\right \rangle \in I$  iff $l(r') = l(v_\mathrm{leaf})$. Substitution is illustrated in Fig. \ref{fig:substitution}.
	\item[-] Adjunction is a tree-insertion operation that allows one to insert an auxiliary tree $\gamma' = \left \langle V',E',r',f'\right \rangle \in A$ at an internal vertex $v_\mathrm{int}$ of a syntactic tree $\gamma = \left \langle V,E,r\right \rangle$. The adjunction operation is defined iff $l(v_\mathrm{int}) = l(r')$. Adjunction takes place in three steps. First, detach the sub-tree $\gamma''=\left \langle V'',E'',v_\mathrm{int}\right \rangle$ starting at the internal node $v_\mathrm{int}$. Subsequently, substitute the foot node $f'$ with the tree $\left \langle V'',E'',v_\mathrm{int}\right \rangle$. This is valid since $l(v_\mathrm{int}) = l(r') = l(f')$. Finally, insert the new syntactic tree in the original tree in place of the the internal vertex $v_\mathrm{int}$. Adjunction is illustrated in Fig. \ref{fig:adjunction}.
\end{itemize}

\subsection{TAG representation of polynomial NARX models}

The discrete-time polynomial NARX model class can be represented as
\begin{equation}
	y_k = \sum_{i=1}^p c_i \prod_{j=0}^{n_u} u_{k-j}^{b_{i,j}} \prod_{m=1}^{n_y} y_{k-m}^{a_{i,m}} + \xi_k, \label{eq:NARX}
\end{equation}
where $p$ is the number of model terms, $u_k, y_k \in \mathbb{R}$ are the input and output at time instant $k$, $\xi_k$ is a white noise process, $c_i$ are the model parameters, and $a_{i,m}, b_{i,j} \in \mathbb{Z}_{\ge0}$ are the exponents for the $m^\text{th}$ output factor and the $j^\text{th}$ input factor in the $i^\text{th}$ model term. In \eqref{eq:NARX}, we can observe a hierarchy in the structure of the Right Hand Side (RHS) of the model expression. This hierarchy can be summarized as follows
\begin{itemize}
	\item[-] a model expression consists of a sum of \textit{terms}, i.e., parts of the expression connected with an addition or subtraction,
	\item[-] each term is a multiplication of \textit{factors}, i.e., parts of the expression connected by multiplication. These factors may be real parameters, input or output signals,
	\item[-] each input and output factor may contain delays.
\end{itemize}
We assign the labels \texttt{expr0, expr1, expr2} for each of the three levels of hierarchy. Furthermore, we use \texttt{op, aff} and \texttt{par} as labels for operators, the affine error term $\xi_k$ and the real coefficients $c_i$ respectively. For convenience, the time index $k$ is dropped. This should not lead to ambiguity since delays will be explicitly denoted by the backward shift operator $q^{-1}$. The TAG for polynomial NARX models is proposed as follows.

\begin{figure}
	\centering
	\includegraphics[width = \linewidth]{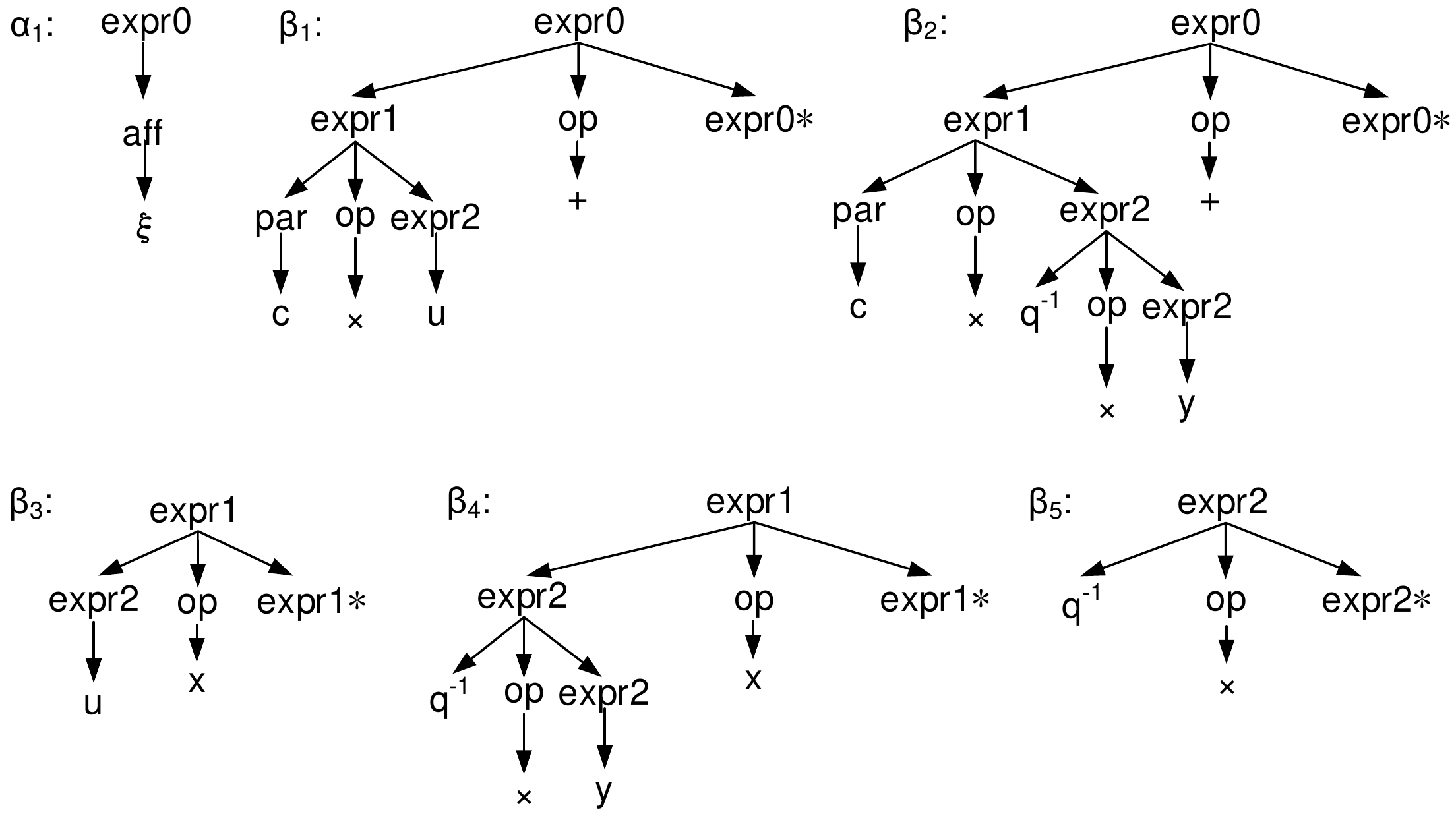}
	\caption{Initial tree $I = \{ \alpha_1 \}$ and auxiliary trees $A = \{ \beta_i \}_{i=1}^5$ of the TAG $G_\mathrm{NARX}$. Symbol $*$ marks the foot node of the auxiliary tree.}
	\label{fig:TAG_NARX}
\end{figure}
		
\begin{proposition}
	The TAG of polynomial NARX models is $G_\mathrm{NARX} = \left \langle N,T,S,I,A \right \rangle$ with
	\begin{itemize}
		\item[-] $N = \{\mathrm{expr0, expr1, expr2, op, aff, par} \}$,
		\item[-] $T = \{ \mathrm{u, y, \xi, c, +, } \times \mathrm{, } q^{-1} \}$,
		\item[-] $S = \{\mathrm{expr0}\}$,
		\item[-] $I = \{\alpha_1\}$, with the initial tree $\alpha_1$ depicted in Fig. \ref{fig:TAG_NARX},
		\item[-] $A = \{\beta_1, \beta_2, \beta_3, \beta_4, \beta_5 \}$, with the auxiliary trees $\{ \beta_i \}_{i=1}^5$ depicted in Fig. \ref{fig:TAG_NARX}.
	\end{itemize}
	The string language $L(G_\mathrm{NARX})$ of grammar $G_\mathrm{NARX}$ is the set of all expressions that can be expressed as the RHS of \eqref{eq:NARX} with finite values of $p, n_u, n_y$.
\end{proposition}
	For brevity, we provide a short sketch of the proof here. The simplest saturated tree that can be generated from $G_\mathrm{NARX}$ is the initial tree $\alpha_1$. The yield of the tree is $\xi$ which corresponds to the simplest model that can be generated:
	\begin{equation}
		y_k = \xi_k.
		\label{eq:simplestExp}
	\end{equation}
	The simplest model \eqref{eq:simplestExp} can be augmented by adjoining the auxiliary trees $A$. Observe that the auxiliary trees have the same structure as the hierarchy in \eqref{eq:NARX}. The adjunction of auxiliary trees $\beta_1,\beta_2$ results in the addition of new input or output terms to the expression, parameterized by a place-holder parameter $c$. The next level of hierarchy is the multiplication with an arbitrary, but finite number of input and output factors. This is achieved by adjunction of auxiliary trees $\beta_3, \beta_4$. Finally, each of these factors may contain an arbitrary, but finite number of delays. This can be achieved by adjoining the auxiliary tree $\beta_5$. %In this way, it can be shown that the yield of all saturated trees generated from the TAG $G_\mathrm{NARX}$ are the RHS expression of \eqref{eq:NARX}.
	In this way, we can conclude that for any NARX model in the from of $\eqref{eq:NARX}$ we can generate a tree satisfying the construction rules of $G_\mathrm{NARX}$ such that the resulting tree is saturated and its yield is the RHS of \eqref{eq:NARX}. Similarly, we can show that any saturated tree in $L_\mathrm{T}(G_\mathrm{NARX})$ has a yield which is in the form of \eqref{eq:NARX}.

\subsection{Discussion}

Representation of dynamical systems using the proposed $G_\mathrm{NARX}$ has some interesting implications both from the system theory and the system identification perspectives. 

\begin{figure}
	\vspace*{0.2cm}
		\centering
	\includegraphics[width = \linewidth]{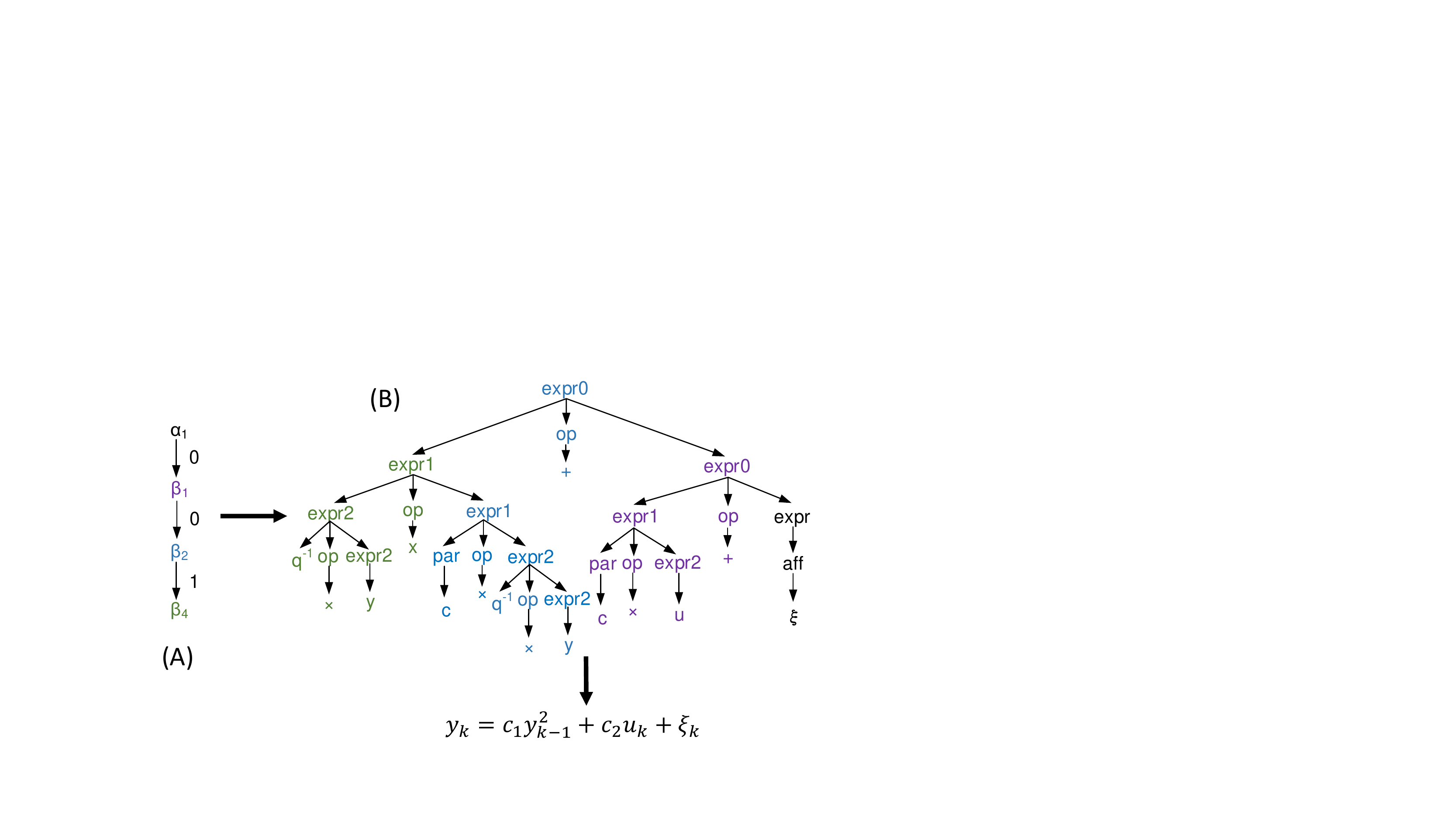}
	\caption{Illustrative example - TAG representation of a NARX model.}
	\label{fig:TAG_NARX_example}
\end{figure}

\subsubsection{Model representation}
TAG provides a new representation for dynamical systems, based on the initial and auxiliary trees used to construct the model. An example is depicted in Fig. \ref{fig:TAG_NARX_example}. In the example, the tree labelled (A) depicts the adjunctions used to construct the model, starting from the initial tree $\alpha_1$. The labels on the edges are the Gorn addresses (see \cite{gorn1965explicit}) of the vertices at which the adjunction takes place. In TAG terminology, this is called the \emph{derivation tree}. In the sequel, derivation trees obtained from the TAG will be used as representations of NARX models. The tree labelled (B) in Fig. \ref{fig:TAG_NARX_example} is the tree generated using $G_\mathrm{NARX}$, and is called the \emph{derived tree}. Notice that the derived tree depicts the model structure on the non-terminal vertices, and the model expression on the leaves.

\subsubsection{Dynamical sub-classes}

Since the set of auxiliary trees $A$ determine the possible ways in which a model may be developed, it should be possible to also restrict the development of models by using a subset of trees in $A$. Indeed, by choosing relevant subsets of auxiliary trees $A$, we can derive the TAG of model classes such as FIR, ARX and truncated Volterra series. For example, choosing the subset $A_1 = \{ \beta_1, \beta_5 \}$ yields the TAG for FIR models, and the subset $A_2 = \{\beta_1, \beta_2, \beta_5 \}$ yields the TAG for ARX models. Similarly, it is reasonable to expect that by adding suitable auxiliary trees, we can obtain a TAG for more flexible model classes such as NARMAX. In a data-driven modelling context, this provides a systematic way to introduce prior knowledge of structural restrictions by the user.

\subsubsection{Ranking of models across various model classes}

In system identification, Occam's Razor principle is often used as a guiding heuristic for model selection between different models. Naturally, a pre-requisite for using this principle is that one should be able to rank the various models in terms of some complexity measure. TAG based representations provide an opportunity to rank models that belong to different model classes. However, such a ranking cannot possibly be 1-dimensional as is usually the case. In order to rank a model, one must take into account not only the dimensions of the TAG representation of the model (for example, the number of vertices in the derivation tree in Fig. \ref{fig:TAG_NARX_example}), but also the specific auxiliary trees used. The latter is necessary since various auxiliary trees add to the complexity of a model differently. For example, adjunction of $\beta_3$ introduces a multiplicative non-linearity, while adjunction of $\beta_1$ introduces a new linear term to a model. This notion can be observed analogously in natural languages, where, in order to assess the meaning of a sentence, one must understand the implications of not only the \emph{syntax} (the precise ordering of the words) but also the \emph{semantics} (the meaning of the individual words).

\section{System Identification using TAG Representations}

\ifComments
\textcolor{red}{
Introduce a method for system identification that is based on TAG representations.
\begin{itemize}
\item Model class and model representation - state that the notion of model class becomes more generalized in the TAG setting. The new model class is related to the generative capacity of the TAG, and goes beyond the traditional borders of model classes.
\item Performance criterion - The traditional choice of performance criterion was based on a ``system in the model class'' assumption. Since the concept of model class is more generalized here, the performance criterion must be revised. Propose multiple criterion (prediction and simulation). Also introduce the idea of pareto-optimality.
\item Numerical approach - introduce TAG3P (Tree Adjoining Grammar Guided Genetic Programming) and LS parameter estimation. Also describe the selection procedure for multi-objective optimization.
\end{itemize}
My concern is that the level of discussion in this section is currently too vague, and on the level of general philosophical commentary. It should be a little more specific, but still somehow compact.
}
\fi

Having introduced TAG representations of dynamical systems, we can now describe a TAG-based approach to identify dynamical systems from measured data using TAG representations. For any system identification problem, three crucial choices have to be made - the model class, the performance criterion and the algorithm and numerical machinery used for model estimation (as per the chosen model class and performance criterion). Each of these choices are introduced and motivated in this Section.

\subsection{Model class and complexity}

Since our aim is to estimate the model structure (and hence, also the model class), the notion of ``model class'' needs to be made more flexible to include a collection of model classes. The proposed $G_{\mathrm{NARX}}$ can be used to generate FIR, ARX, truncated Volterra series or polynomial NARX models. Rather than selecting a set of model classes, we use the notion of ``model class'' to describe the generative capacity of the grammar. In this paper, the chosen model class is the set of all models that can be generated by $G_{\mathrm{NARX}}$. It should be noted that, while the TAG remains a user-choice, it is not as critical as the choice of a specific model class since multiple model classes can be generated by the same TAG.
%
%
%Using TAG, we have proposed a model representation that incorporates the model structure in the representation. Furthermore, we have also proposed a TAG for polynomial NARX models that incorporates the structure of FIR, ARX and truncated Volterra series. Hence, for a TAG-based approach to system identification, the choice of model structure can be replaced by a choice of TAG. In this paper, the scope of identification is restricted to the TAG $G_\mathrm{NARX}$.

The proposed identification method makes use of heuristic algorithms in order to automatically explore models of different complexities. This further alleviates the need for a user to pre-specify the desired model complexity. While the model complexity can be any user-defined measure, in this paper, the complexity of a model generated by $G_\mathrm{NARX}$ is measured in terms of the number of parameters in the model.

TAG provides two new representations of dynamical systems (i.e. derivation trees and derived trees) in addition to the well-known symbolic representation (such as \eqref{eq:NARX}). In the proposed identification approach, operations related to the search of model structure will be based on the derivation tree representation, and operations related to parameter optimization will use the symbolic representation \eqref{eq:NARX}.

\subsection{Performance criterion}

Traditionally, the 1-step-ahead prediction error has been favourably used in the identification of parameters of parametric models such as \eqref{eq:NARX} because of its statistical interpretation. If the to-be-identified system belongs to the model class and if the noise process is zero-mean and white, then the model that minimizes the mean squared prediction error is also the MAP estimate. For several model classes, including \eqref{eq:NARX}, parameter estimation using the 1-step ahead prediction reduces to a LS (Least Squares) problem, which is numerically efficient to solve. However, while it might be possible to generate from the chosen TAG, a model structure that captures the dynamic structure of the original system, it is certainly also possible to generate model structures that are under-parameterized. In such a setting, the use of prediction error alone is not justifiable. 

As an extreme alternative to the one-step-ahead prediction error, we can consider simulation error, which is often interpreted as $\infty$-step-ahead prediction error. It has been observed in \cite{piroddi2003identification} and \cite{aguirre2010prediction} that simulation error is more sensitive to model errors in non-linear systems than prediction error. However, estimating model parameters using simulation error leads to a non-convex optimization problem, in general.

Due to the complimentary benefits and pitfalls, we propose the following performance criterion:
\begin{itemize}
	\item[-] Use 1-step-ahead prediction error to estimate parameters of a given stochastic model,
	\item[-] Use both 1-step-ahead prediction error and simulation error, in a multi-objective optimization framework, to measure the performance of a model, i.e., both model structure and model parameters.
\end{itemize}

Using a non-linear stochastic model for simulation is a non-trivial task. The common approach to simulaiton is to set the noise contributions to 0. However, it has been shown in \cite{khandelwal2018on} that this leads to biased simulation models in the case of non-linear systems. An approach to compute simulation models from stochastic models is proposed in \cite{khandelwal2018on}, and is used here.

\subsection{Algorithm}

This section describes the algorithm for estimation of model structure and model parameters. TAG-based representations and Genetic Programming (GP) are used to solve the problems of estimation model structure. GP uses tree-like representations of solutions, and hence TAG-based representations can also be used. Moreover, since model structure estimation is a combinatorial problem for which no systematic solution exists, it is reasonable to rely on heuristic solution approaches. GP provides a set of biologically-inspired heuristics that have yielded competitive results in multiple domains of science and engineering (for e.g., \cite{arias2012multiobjective}). GP also provides a numerical platform to solve the multi-objective optimization problem using the notion of non-dominated solutions and pareto-optimality \cite{arias2012multiobjective}. For these reasons, GP is a natural choice for the estimation of model structure. An overview of the algorithm is sketched in Algorithm \ref{alg:MOO}.

\begin{algorithm}
	\caption{Multi-objective optimization using TAG3P and LS}
	\label{alg:MOO}
	\begin{algorithmic}[1]
		\Require population size $M>0$, number of iterations $L>0$, grammar $G$, probabilities of crossover and mutation $p_c,p_m$
%		\State Initialize population $X^{(0)}$ of size $M$ \Comment{See \cite{koza1992genetic}}
%		\State Estimate model parameters in $X^{(0)}$ using LS %\Comment{See \textcolor{red}{cite}}
%		\State Compute multi-objective fitness of models in $X^{(0)}$
		\State Initialize population $X^{(0)}$, $l=0$, $X^{(-1)}=\{ \}$ \Comment{See \cite{koza1992genetic}}
%		\State $l \leftarrow 1$
		\Repeat
			\State Estimate model parameters in $X^{(l)}$ using LS \Comment{See \cite{khandelwal2018on}} \label{step:parEst}
			\State Compute multi-objective fitness of models in $X^{(l)}$ \label{step:multObj}
			\State Perform non-dominated sorting of populations $X^{(l-1)}$ and $X^{(l)}$ \Comment{See \cite{deb2002fast}}
			\State $X^{(l)} \leftarrow$ first $M$ individuals of the sorted combined population.
			\State Propose new population $X^{(l+1)}$ using crossover and mutation \Comment{See \cite{hoai2003tree}}
			\State $l \leftarrow l+1$
		\Until{$l \le L+1$}
		\Return $X^{(L)}$
	\end{algorithmic}
\end{algorithm}

\subsubsection{Genetic Programming}
%GP is an optimization method based on meta-heuristics inspired from the theory of evolution.
GP is an iterative scheme that develops and propagates a set of solutions (the population) iteratively. In each iteration, a new set of population is proposed by using \emph{genetic operators} such as \emph{crossover} and \emph{mutation}. Subsequently, a \emph{selection} scheme is used to select a number of solutions from the existing and the newly proposed solutions, based on a user-defined performance measure of the models. The selected models are propagated to the next iteration, and this process is repeated for a fixed number of maximum iterations. More information on GP can be found in \cite{koza1992genetic}.

In the proposed algorithm, we use a variant of GP called Tree Adjoining Grammar-Guided GP (TAG3P) \cite{hoai2003tree}, that was developed for TAG formulations. Hence, the genetic operations of crossover and mutation are adapted for derivation tree representation of the model. The search scheme is initialized with $M$ randomly generated derivation trees from $G_\mathrm{NARX}$. In each iteration, a non-dominated sorting and selection algorithm, proposed in \cite{deb2002fast}, is used to select and propagate pareto-optimal solutions. Other hyper-parameters involved are the number of iterations $L$, the probability for crossover $p_c$ and the probability of mutation $p_m$. A high probability of crossover ensures that sub-structures of models that achieve better performance are transferred to other models in the population. A high probability of mutation allows for more frequent random explorations in the space of models. As there is no guarantee of convergence, there is no systematic choice of the maximum number of iterations $L$. One possible approach is to select a conservatively large $L$, and terminate the algorithm when a desired level of performance has been achieved. %Each derivation tree in a generation is denoted by $x_i^j$, where $i \in [1,M]$ is the index of the model in the generation, $j \in [0,L]$ is the index of the generation, and $L$ is the maximum number of generations. Subsequently, crossover operator is applied to random pairs ()

\subsubsection{Parameter estimation}
In each iteration of GP, crossover and mutation are used to propose a new set of solutions from the solutions of the previous iterations. The new solutions, however, cannot be used in the non-dominated sorting algorithm due to the presence of yet-to-be-determined model parameters. Since TAGs may generate models that belong to different model classes, the parameter estimation approach should be robust enough to deal with variability of model structures. Black-box numerical optimization methods such as CMA-ES (see \cite{hansen2001completely}) can be used to estimate model parameters in a linear or non-linear setting. An alternative, more efficient approach would be to make use of one out of a collection of parameter estimation algorithms, depending on the model structure generated by the grammar. The use of TAG makes it possible to systematically infer the model structure, and hence the appropriate optimization algorithm.

Since the scope of this paper is limited to polynomial NARX models, the parameter estimation problem can be solved efficiently. Minimization of the sum-of-squares of the prediction error of \eqref{eq:NARX} leads to a quadratic cost function that can be solved using LS \cite{Ljung1999}.% On the other hand, minimization of the simulation error, or any combination of simulation and prediction error, typically leads to non-convex optimization problem.

%\begin{remark}
%It should be noted that the choice of LS estimation is indeed dependent on the grammar $G_{\mathrm{NARX}}$. This dependency can be alleviated by the use of black-box non-linear parameter estimation methods such as CMA-ES \cite{hansen2001completely}. However, this comes at the cost of significantly higher computation times. An alternative, more efficient approach would be to make use of one out of a collection of parameter estimation algorithms, depending on the model structure generated by the grammar. The use of TAG makes it possible to programmatically decide the model structure, and hence the appropriate optimization algorithm.
%\end{remark}

\section{Identification results}

\begin{figure}
	\centering
	\includegraphics[width = 0.8\linewidth]{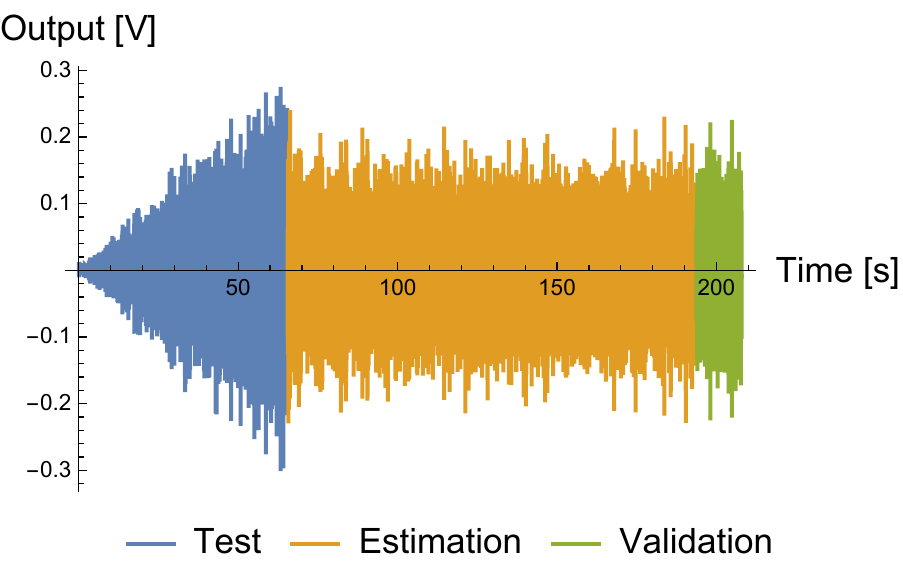}
	\caption{Silverbox measured data.}
	\label{fig:silverboxData}
\end{figure}

To illustrate the proposed TAG-based identification procedure, we use the benchmark Silverbox data-set proposed in \cite{wigren2013three}. The silverbox system in an electronic implementation of a mass-spring-damper system with a non-linear spring. The data-set, measured at a sampling rate of 610.35 Hz, consists of 2 parts. The excitation signal used in the first 40000 samples is a low-pass filtered Gaussian noise signal of cut-off frequency 200 Hz and a linearly increasing amplitude ranging from 0 to a maximum value of about 0.3 V. The data-set is plotted in Fig. \ref{fig:silverboxData}.  The excitation signal in the second part of the data-set consists of 10 realizations of a random odd multi-sine signal (see \cite{wigren2013three} for details). We use the first 9 realizations of the multi-sine data-set as \emph{estimation data-set} to estimate model parameters in Step \ref{step:parEst}, and the last realization as \emph{validation data-set} to evaluate the multi-objective fitness of the proposed models in Step \ref{step:multObj}. Furthermore, the Gaussian excitation signal associated part of the data-set is used as a \emph{test data-set} that is independent of the remaining data used during the identification procedure.

\begin{table}[tb]
\caption{Algorithm hyper-parameters}
\label{tab:hyperparameters}
\centering
\begin{tabular}{l|l}
\hline
\textbf{Hyper-parameter  }              & \textbf{Value }            \\ \hline
Population Size $M$            & 100               \\
Maximum GP iterations $L$      & 150               \\
Maximum adjunctions            & 150               \\
Probability of crossover $p_c$ & 1                 \\
Probability of mutation $p_m$  & 0.8               \\
Grammar                        & $G_\mathrm{NARX}$
\end{tabular}
\vspace*{-0.6cm}
\end{table}

\begin{figure}
		\begin{subfigure}[t]{\linewidth}
			\centering
			\includegraphics[scale = 0.75]{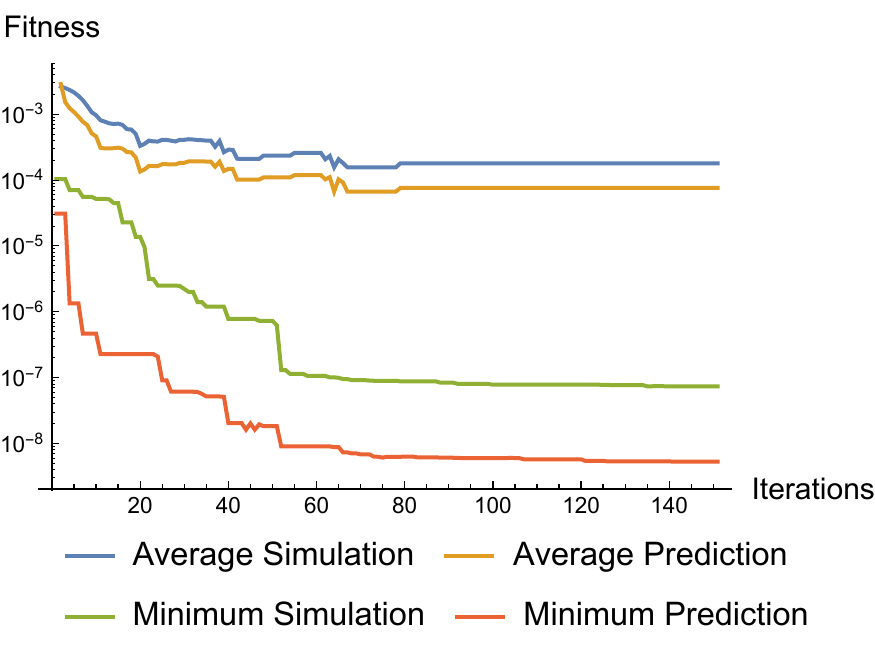}
			\caption{Evolution of performance measures over iterations.}
			\label{fig:silverboxEvo}
		\end{subfigure}
		\\
%		\begin{subfigure}[t]{\linewidth}
%			\centering
%			\includegraphics[scale = 0.75]{Images/silverBoxSimPredPareto.eps}
%			\caption{Pareto-Fronts: complexity vs. simulation and complexity vs. prediction.}
%			\label{fig:silverboxCombinedPareto}
%		\end{subfigure}
		\begin{subfigure}[t]{0.45\linewidth}
			\includegraphics[scale = 0.75]{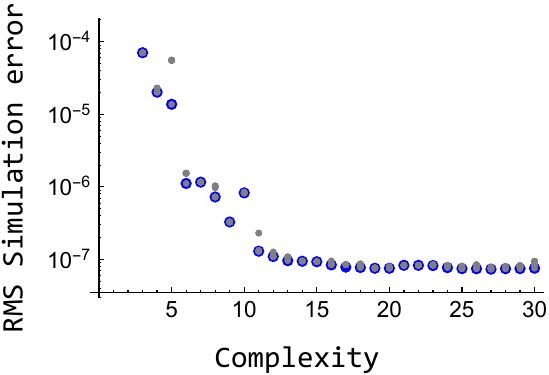}
			\caption{Pareto front - simulation vs complexity.}
			\label{fig:silverboxSim}
		\end{subfigure}
		~~%
		\begin{subfigure}[t]{0.45\linewidth}
			\includegraphics[scale = 0.75]{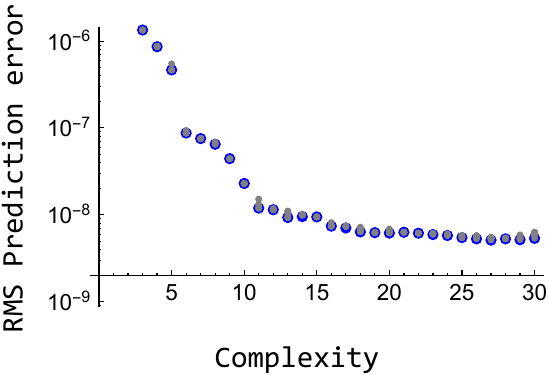}
			\caption{Pareto front - prediction vs complexity.}
			\label{fig:silverboxPred}
		\end{subfigure}
		\caption{Illustration of numerical results.}
		\label{fig:silverboxOverview}
\end{figure}

The hyper-parameters used in the algorithm are given in Table \ref{tab:hyperparameters}. As there are no guarantees of convergence, the hyper-parameters are chosen conservatively. It should be noted that no other information specific to the benchmark example was incorporated into the identification procedure. An illustration of the results obtained are plotted in Fig. \ref{fig:silverboxOverview}. To plot these figures, the complexity measure was chosen as the number of real parameters in the model. Fig. \ref{fig:silverboxEvo} depicts the evolution of the performance measures (average and minimum over all models in the population) with respect to GP iterations. Fig. \ref{fig:silverboxSim} and \ref{fig:silverboxPred} depict 2-D projections of the final pareto-front obtained from the identification procedure, the bold dots indicate the pareto-front of the best models found and the smaller dots indicate some of the sub-optimal models found during the procedure. It should be noted that the bold dots in both Fig. \ref{fig:silverboxSim} and \ref{fig:silverboxPred} correspond to the same models for each level of complexity.

\begin{figure}
		\vspace*{0.2cm}
		\begin{subfigure}[t]{\linewidth}
			\centering
			\includegraphics[scale = 0.65]{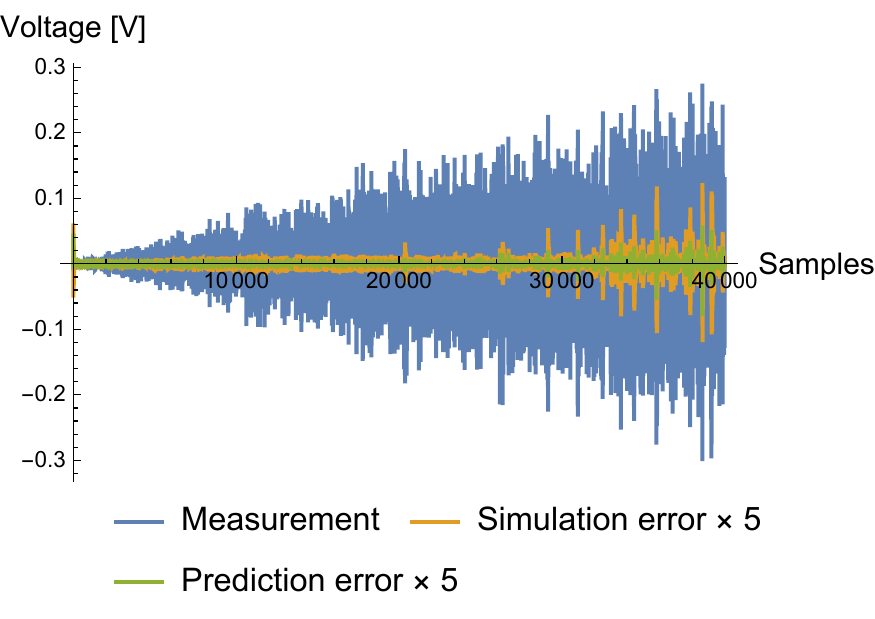}
			\caption{Evaluation of model $M_1$ on test data-set.}
			\label{fig:silverboxM1}
		\end{subfigure}
		\\
		\begin{subfigure}[t]{\linewidth}
			\centering
			\includegraphics[scale = 0.65]{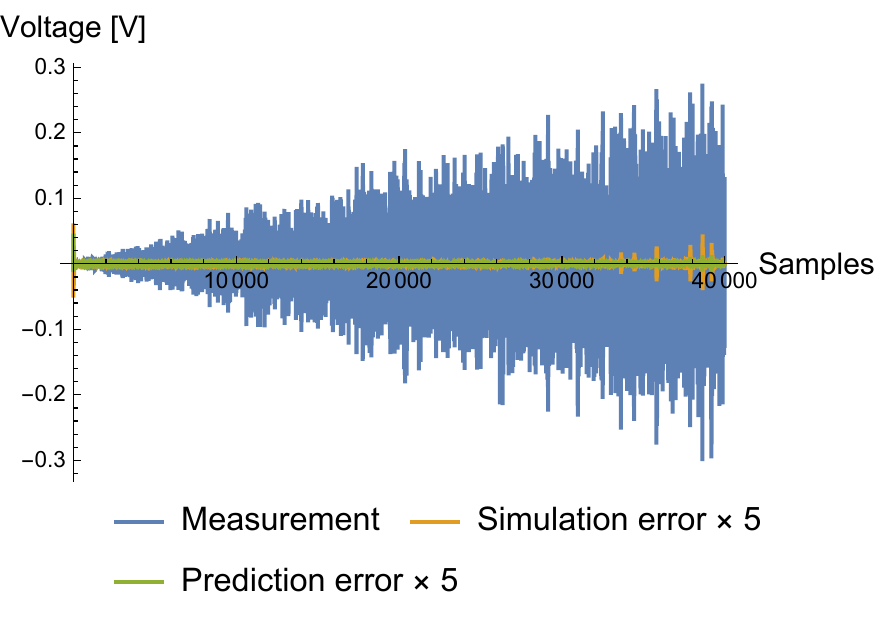}
			\caption{Evaluation on model $M_2$ on test data-set.}
			\label{fig:silverboxM2}
		\end{subfigure}
		\caption{Illustration of numerical results.}
		\label{fig:silverboxOverviewModelEval}
\end{figure}

From Fig. \ref{fig:silverboxSim} and \ref{fig:silverboxPred} we observe that beyond the complexity level of 6 (i.e. 6 model parameters), the improvement in both performance measures is only gradual. Furthermore, the model that achieves the best performance in both measures contains 27 parameters. For the sake of analysis and comparison, we select two models from the pareto-front - $M_1$ corresponding to the best model containing 6 parameters, and $M_2$ corresponding the best model with 19 parameters. Model $M_1$ is described by the following equation
\begin{multline}
	y_k = 0.3694 u_{k-1} + 0.0467 u_{k} + 0.1024 y_{k-3} - 1.0939 y_{k-2} \\
	\quad + 1.5809 y_{k-1} - 1.3923 y_{k-1}^3 + \xi_k.
\end{multline}
The structure of $M_1$ is the same as the structure of the dynamics of the ideal circuit as reported in \cite{schoukens2003fast}. However, in \cite{schoukens2003fast}, it was also reported that the realization of the electronic circuit was non-ideal, which explains the steady improvement in the pareto-fronts in Fig. \ref{fig:silverboxOverview} beyond the model complexity of 6. Fig. \ref{fig:silverboxOverviewModelEval} depicts the prediction and simulation errors (scaled up by a factor of 5) of models $M_1$ and $M_2$. It can be observed in Fig. \ref{fig:silverboxM2} that $M_2$ performs adequately well also in the latter half of the test data-set where the model is required to extrapolate (since the magnitude of the input is greater than that in the estimation and validation set). The performance metrics of $M_1$ and $M_2$ are given in Table \ref{tab:results}. The performance of the proposed identification procedure is comparable to state-of-the-art non-linear system identification methods. For instance, the results obtained by PNLSS identification \cite{marconato2012identification} are also provided in Table \ref{tab:results} for comparison. While the PNLSS method produces a single model, in this case with 37 parameters, the proposed approach provides a pareto-front of models. This enables the user to choose the performance-complexity trade-off \textit{a-posteriori}. Furthermore, in the proposed method, the user is not required to make any critical choices, while in the case of PLNSS, the user must make well-informed decisions, such as the order of non-linearity considered and the initialization method.

\begin{table}[]
\vspace*{-0.2cm}
\caption{Numerical results achieved and compared with literature. The blank cells correspond to values not reported in the literature.}
\label{tab:results}
\begin{tabular}{l|c|c|c|c}
\hline
\multicolumn{1}{c|}{\begin{tabular}[c]{@{}c@{}}Identified\\ model\end{tabular}} & \begin{tabular}[c]{@{}c@{}}Test RMS\\ simulation\\ (mV)\end{tabular} & \begin{tabular}[c]{@{}c@{}}Test RMS\\ prediction\\ (mV)\end{tabular} & \begin{tabular}[c]{@{}c@{}}Val. RMS\\ simulation\\ (mV)\end{tabular} & \begin{tabular}[c]{@{}c@{}}Val. RMS\\ prediction\\ (mV)\end{tabular} \\ \hline
$M_1$                                                                           & 1.8046                                                               & 0.4525                                                               & 1.235                                                                & 0.2941                                                               \\
$M_2$                                                                           & 0.4196                                                               & 0.1017                                                               & 0.2698                                                               & 0.0731                                                               \\
\begin{tabular}[c]{@{}l@{}}Best Linear\\ Approx. \cite{marconato2012identification}\end{tabular}             & 13.5                                                                 & -                                                                    & 6.9                                                                  & -                                                                    \\
PNLSS \cite{marconato2012identification}                                                                          & 0.26                                                                 & -                                                                    & -                                                                    & -                                                                   
\end{tabular}
\vspace*{-0.7cm}
\end{table}
\ifComments
\textcolor{red}{
\begin{itemize}
\item Describe the silverbox example.
\item Specify the hyper-parameters of the identification procedure.
\item Discuss results and compare with literature.
\end{itemize}
}
\fi
\section{CONCLUSIONS}

We proposed a new identification procedure that uses TAG-based representations to identify both the structure and parameters of a model. The proposed framework is formulated on auto-regressive forms ranging from linear to (non-linear) polynomial models. The numerical illustration demonstrates that almost no prior information of the system dynamics is required during the identification procedure itself. In fact, the model structure of the true dynamical system was correctly estimated by the method with no prior knowledge. The illustration also demonstrates the potential of achieving automation of system identification using the proposed procedure. %Extending the current formulation to more general non-linear system descriptions will be a subject of future research.
TAG-based representations are vital to this approach as they provide a mechanism to isolate the numerical algorithm from the choice of model class, thereby allowing systematic exploration of model structures among different model classes. While the scope of the paper was restricted to illustrate the idea efficiently, it is certainly possible to extend the idea to general non-linear system descriptions and beyond.

\addtolength{\textheight}{-6cm}   % This command serves to balance the column lengths

%%%%%%%%%%%%%%%%%%%%%%%%%%%%%%%%%%%%%%%%%%%%%%%%%%%%%%%%%%%%%%%%%%%%%%%%%%%%%%%%

%%%%%%%%%%%%%%%%%%%%%%%%%%%%%%%%%%%%%%%%%%%%%%%%%%%%%%%%%%%%%%%%%%%%%%%%%%%%%%%%

%%%%%%%%%%%%%%%%%%%%%%%%%%%%%%%%%%%%%%%%%%%%%%%%%%%%%%%%%%%%%%%%%%%%%%%%%%%%%%%%
%\section*{APPENDIX}
%\section*{ACKNOWLEDGMENT}
%
%The preferred spelling of the word �acknowledgment� in America is without an �e� after the �g�. Avoid the stilted expression, �One of us (R. B. G.) thanks . . .�  Instead, try �R. B. G. thanks�. Put sponsor acknowledgments in the unnumbered footnote on the first page.

%%%%%%%%%%%%%%%%%%%%%%%%%%%%%%%%%%%%%%%%%%%%%%%%%%%%%%%%%%%%%%%%%%%%%%%%%%%%%%%%

%References are important to the reader; therefore, each citation must be complete and correct. If at all possible, references should be commonly available publications.

\bibliographystyle{IEEEtran}

\bibliography{grammar_NARX}

\end{document}